\documentclass{PoS}

\usepackage{amsmath, amssymb, amsfonts, bbold, subfigure, dsfont}

\newcommand{\secret}[1]{}
\newcommand{\ddd}{\displaystyle}
\newcommand{\bra}[1]{\langle #1|}
\newcommand{\ket}[1]{|#1\rangle}
\newcommand{\vac}{|0\rangle}

\newcommand{\comm}[2]{\left[#1, #2\right]}

\newcommand{\comment}[1]{}

\newcommand{\lr}[1]{ \left( #1 \right) }

\newcommand{\vev}[1]{ \langle \, #1 \, \rangle }

\title{Topological and magnetic properties of the QCD vacuum probed by overlap fermions
\thanks{The calculations were done at the GSI Batch Farm (Darmstadt) and on the BIRD farm at DESY}}

\ShortTitle{Topological and magnetic properties of the QCD vacuum probed by overlap fermions}

\author{V.V. Braguta\\
        IHEP, Protvino, Moscow oblast, 142284 Russia\\
	ITEP, B. Cheremushkinskaya str. 25, Moscow, 117218 Russia\\
        E-mail: \email{braguta@itep.ru}}

\author{P.V. Buividovich\\
        ITP, University of Regensburg, Universit\"atsstrasse 31, D-93053 Regensburg, Germany\\
        E-mail: \email{pavel.buividovich@physik.uni-regensburg.de}}

\author{\speaker{T. Kalaydzhyan}\\
        DESY Hamburg, Theory Group, Notkestrasse 85, D-22607 Hamburg, Germany\\
        E-mail: \email{tigran.kalaydzhyan@desy.de}}

\author{M.I. Polikarpov\\
        ITEP, B. Cheremushkinskaya str. 25, Moscow, 117218 Russia\\
        E-mail: \email{polykarp@itep.ru}}

\abstract{We study some of the local CP-odd and magnetic properties of the non-Abelian vacuum with use of overlap fermions within the quenched lattice gauge theory. Among these properties are the following:
  inhomogeneous spatial distribution of the topological charge density (chirality for massless fermions) in SU(2) gluodynamics (for uncooled gauge configurations the chirality is localized on low-dimensional defects with $d=2 .. 3$, while a sequence of cooling steps gives rise to four-dimensional instantons and hence a four-dimensional structure of the chirality distribution); finite local fluctuations of the chirality growing with the strength of an external magnetic field; magnetization and susceptibility of the QCD vacuum in SU(3) theory; magnetic catalysis of the chiral symmetry breaking, and the electric conductivity of the QCD vacuum in strong magnetic fields.}

\FullConference{Xth Quark Confinement and the Hadron Spectrum\\
                 8--12 October 2012\\
                 TUM Campus Garching, Munich, Germany}

\begin{document}

\section{Introduction}
\begin{floatingfigure}[r]
\label{distribution}
\includegraphics[angle=0, width=5cm]{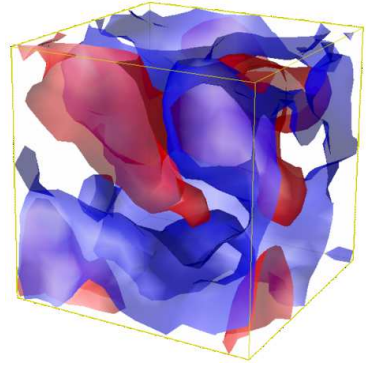}\hspace{0cm}
\caption{Isosurfaces of the chirality distribution, see \cite{Buividovich:2011cv} for details.}
\end{floatingfigure}
Recently both aspects of our studies, properties of the QCD vacuum in strong magnetic fields and its non-trivial
topological structure, attracted much attention in light of ongoing heavy-ion experiments. The magnitudes of the magnetic fields possibly created there are of order of $B \sim m_\pi^2$, which may change the critical temperature of the chiral transition, several electromagnetic properties of QCD and also give rise to new phenomenological effects potentially measurable experimentally. For the origin and properties of the magnetic fields in heavy-ion collisions see the review \cite{Tuchin:2013ie}. 

A non-trivial gluonic background may create an imbalance between numbers of left- and right-handed quarks and lead to the local strong CP-violation \cite{Kharzeev, CME}. A typical spatial distribution of such imbalance is shown in Fig.~\ref{distribution}, where the colors denote some fixed positive (red) and negative (blue) values of chirality. 
It is remarkable, that for uncooled gauge field configurations the distribution is irregular and, as we show later, does not follow the instanton pattern, but forms low-dimensional structures.

\section{Magnetic-field-induced effects}\label{first}
\subsection{Technical details}

We use a quenched $SU(3)$ lattice gauge theory with the tadpole-improved
L\"uscher-Weisz action \cite{Luscher}. To generate statistically
independent gauge field configurations we implement the Cabibbo-Marinari heat bath
algorithm. The lattice size is $14^4$, and lattice spacing $a=0.105fm$. All
observables we discuss later have a similar structure: $\langle\bar\Psi
\mathcal{O} \Psi\rangle$ for VEV of a single quantity or $\langle\bar\Psi
\mathcal{O}_1 \Psi \,\,\, \bar\Psi \mathcal{O}_2 \Psi\rangle$ for dispersions
or correlators. Here $\mathcal{O}$, $\mathcal{O}_1$, $\mathcal{O}_2$ are some
operators in spinor and color space. These expectation values can be expressed
through the sum over $M$ low-lying\footnote{We believe that the IR quantities
are insensitive to the UV cutoff realized by selecting some finite number of
the eigenmodes \cite{Hasenfratz}} but non-zero eigenvalues $i\lambda_k$ of the
chirally invariant Dirac operator $D$ (Neuberger's overlap Dirac
operator \cite{Neuberger}):
\begin{align}
\label{single} \langle\bar\Psi \mathcal{O} \Psi\rangle =
\sum\limits_{|k|<M}\frac{\psi_k^{\dag}\mathcal{O}\psi_k}{i\lambda_k + m},  \qquad  \mathrm{where} \qquad D \psi_k = i \lambda_k \psi_k,
\end{align}
and
\begin{align}
\label{double}\langle\bar\Psi \mathcal{O}_1 \Psi \,\,\, \bar\Psi \mathcal{O}_2
\Psi\rangle = \sum\limits_{k,
p}\frac{\bra{k}\mathcal{O}_1\ket{k}\bra{p}\mathcal{O}_2\ket{p}-\bra{p}\mathcal{O}_1\ket{k}\bra{k}\mathcal{O}_2\ket{p}}{(i\lambda_k+m)(i\lambda_p+m)},
\end{align}
where all spinor and color indices are contracted and we drop them for
simplicity. The uniform magnetic
field $F_{12}=B_3\equiv B$ is introduced as described
in \cite{Buividovich_condensate}. To perform calculations in the chiral limit
we calculated the expression (\ref{single}) or (\ref{double}) for a
non-zero $m$ and averaged it over all configurations of the gauge fields. Then
we repeated the procedure for other quark masses $m$ and extrapolated the VEV
to $m\rightarrow 0$ limit.
Simulation details for the Sections \ref{second} and \ref{third} are similar, but for two colors, and are described in the mentioned there papers.

\subsection{Chiral condensate}
\label{condensatesection}
\begin{figure}[t]
     \centering
     \subfigure[\label{sigma}]
     {\includegraphics[angle=-90, width=7.5cm]{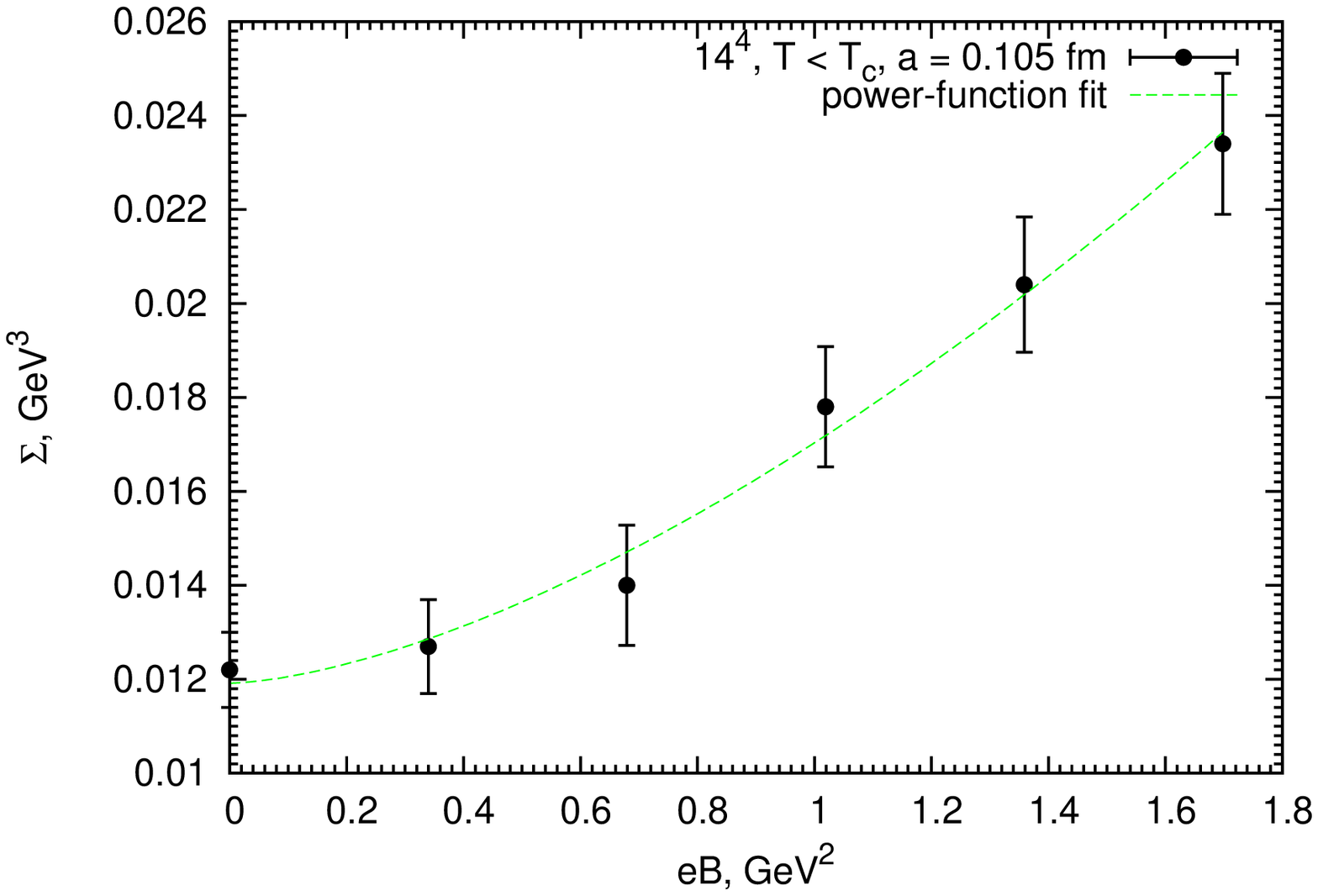}}\hspace{0cm}
     \centering
     \subfigure[\label{s}]
     {\includegraphics[angle=-90, width=7.5cm]{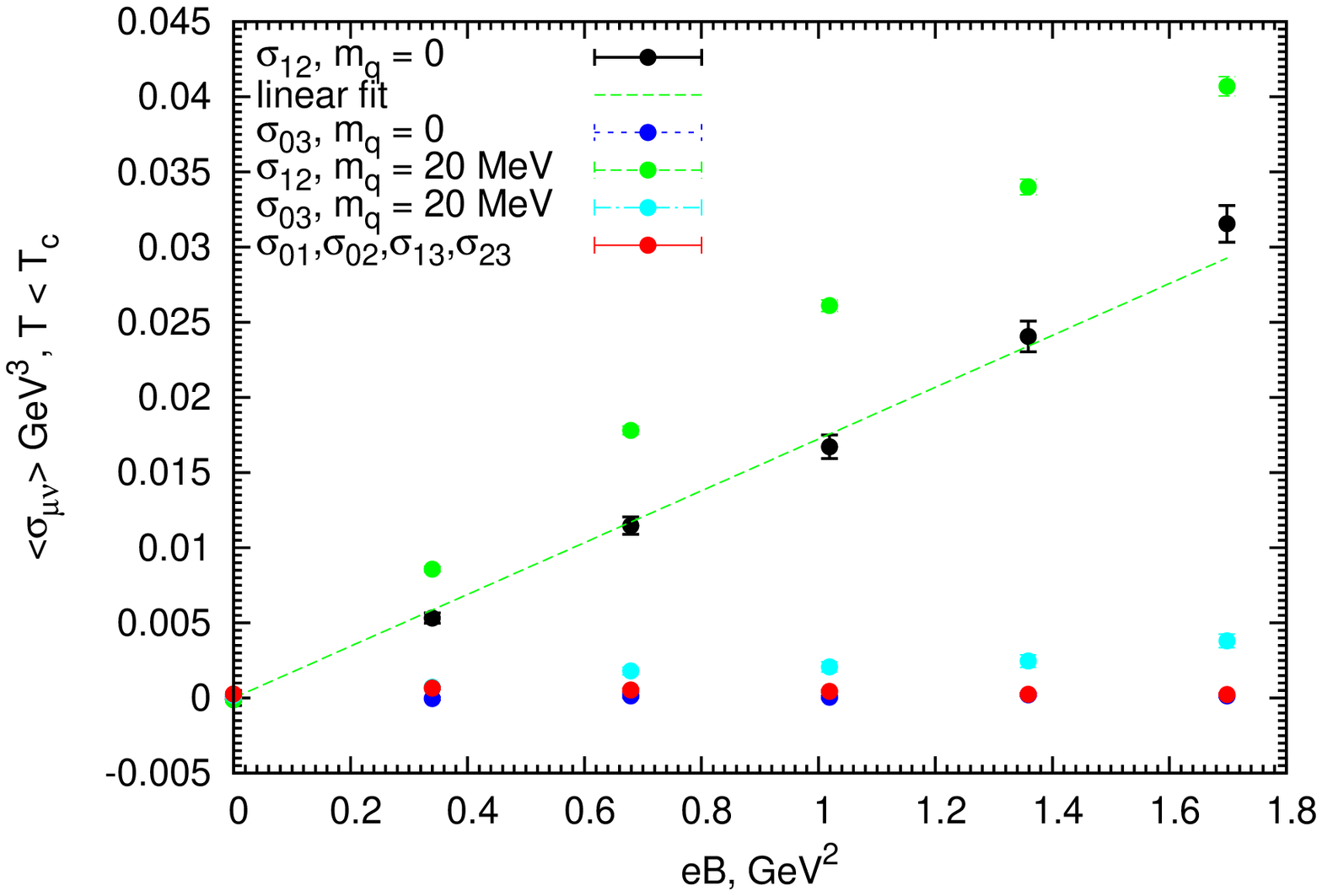}}\hspace{0cm}
\caption{\label{condensateplots} Chiral condensate (a) and magnetization/polarization $\langle\bar \Psi
\sigma_{\alpha\beta} \Psi\rangle$ (b).}
\end{figure}
\noindent In this section we present our results for the chiral condensate ($\mathcal{O} = \mathds{1}$),
\begin{align}
\label{condensate}
 \Sigma \equiv -\bra{0} \bar \Psi \Psi \vac,
\end{align}
as a function of the magnetic field $B$. The result is shown in Fig.~\ref{sigma}. The general tendency for $\Sigma$ to grow with $B$ has been obtained in various models (see \cite{Shovkovy} for a review) and 
usually referred to the magnetic catalysis of the chiral symmetry breaking\footnote{A non-monotonic behavior in the vicinity of $T_c$ has been recently observed
\cite{Bali:2011qj, Bali:2012zg} and named as \textit{inverse magnetic catalysis}.}. 
We perform the fit of the results by the following function
\begin{align}
 \Sigma^{fit}(B) = \Sigma_0\left[ 1 + \left(\frac{eB}{\Lambda_B^2}\right)^{\nu}\right],
\end{align}
where $\Sigma_0 \equiv \Sigma(0)$. The obtained fitting parameters are
\begin{align}
 \Sigma_0 = \left[(228 \pm 3) \,\mathrm{MeV}\right]^3, \qquad \Lambda_B = \left( 1.31 \pm 0.04 \right)\,\mathrm{GeV}, \qquad \nu = 1.57 \pm 0.23 \,.
\end{align}
The value of the condensate in absence of the magnetic field $\Sigma(0)$ and the exponent $\nu$ have reasonable values, compare with e.g. \cite{Colangelo,D'Elia:2011zu}.

\subsection{Magnetization and susceptibility}
\label{magnetizationsection}

\noindent In this section we calculate the quantity
\begin{align}
\label{sigma_def}
 \langle \bar \Psi \sigma_{\alpha\beta} \Psi \rangle = \chi(F) \langle\bar\Psi\Psi\rangle q F_{\alpha\beta},
\end{align}
where $\sigma_{\alpha\beta}\equiv\ddd\frac{1}{2 i}
\comm{\gamma_{\alpha}}{\gamma_{\beta}}$ and $\chi(F)$ is a coefficient of
proportionality (susceptibility), which depends on the field strength.

This quantity was introduced in \cite{Ioffe} and can be used to estimate the
spin polarization of the quarks in external magnetic field. The magnetization
can be described by the dimensionless quantity $\label{mu_def} \mu  = \chi
\cdot qB$, so that
\begin{align}
\label{mu_in_eq} \langle\bar\Psi\sigma_{12}\Psi\rangle = \mu
\langle\bar\Psi\Psi\rangle \,.
\end{align}

The expectation value (\ref{sigma_def}) can be calculated on the lattice by
(\ref{single}) with $\mathcal{O} = \sigma_{\alpha\beta}$. The result is shown
in Fig.~\ref{s}. We can see, that the 12-component grows linearly with the field, which
agrees with \cite{Ioffe}. This allows us to find the chiral susceptibility
$\chi(0)\equiv\chi_0$ from the slope of the curve. After a linear approximation
$\langle\bar\Psi\sigma_{12}\Psi\rangle = \Omega^{fit} eB$, where\footnote{in
our simulation we calculate the magnetization of the d-quark condensate, thus
$q=\left|-\frac{e}{3}\right|$}
\begin{align}
\Omega^{fit} \equiv -\frac{1}{3}\chi_0^{fit}\Sigma_0 \,,
\end{align}
we obtain $\Omega^{fit} = (172.3 \pm 0.5) \,\mathrm{MeV}$ and
\begin{align}
\chi_0^{fit}= -4.24 \pm 0.18\, \mathrm{GeV}^{-2}\,.
\end{align}

This value fits well into the range of present theoretical estimations (see \cite{Frasca:2011zn} for a review and \cite{Bali:2012jv} for recent results).
We also have to mention a well known analytic result obtained by the OPE combined with the idea of pion
dominance \cite{Vainshtein}, supported by two holographic ones \cite{SusGorsky, SusSon}, but giving a too high value
comparing to our results. This disagreement seems to be an important puzzle to solve, because it will lead to a better understanding of the pion dominance assumption and the large $N_c$ limit.

\subsection{Evidences of the chiral magnetic effect}

\begin{figure}[t]
     \centering
     \subfigure[\label{r52IR}]
     {\includegraphics[angle=-90, width=7.5cm]{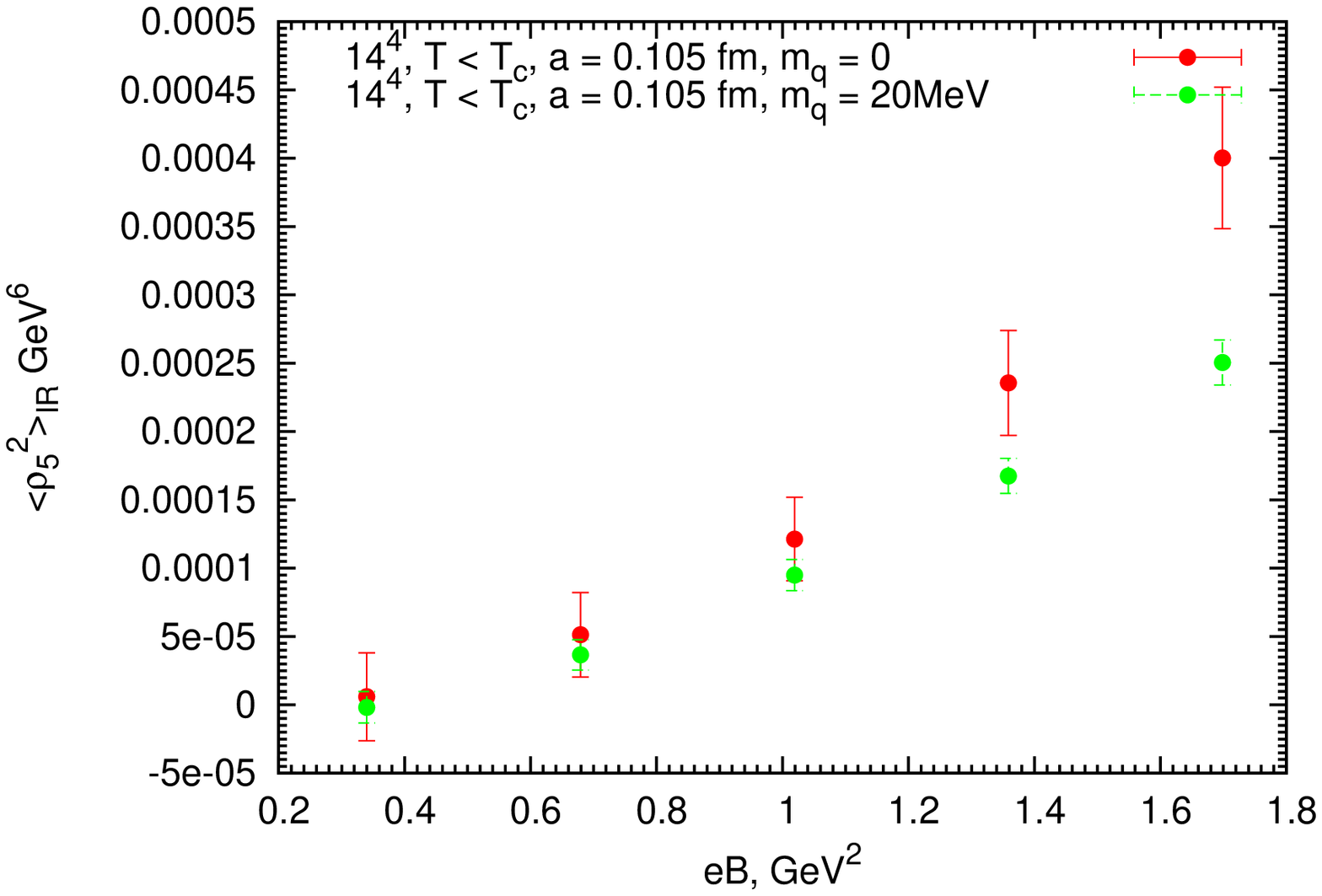}}\hspace{0cm}
     \centering
     \subfigure[\label{j2IR}]
     {\includegraphics[angle=-90, width=7.5cm]{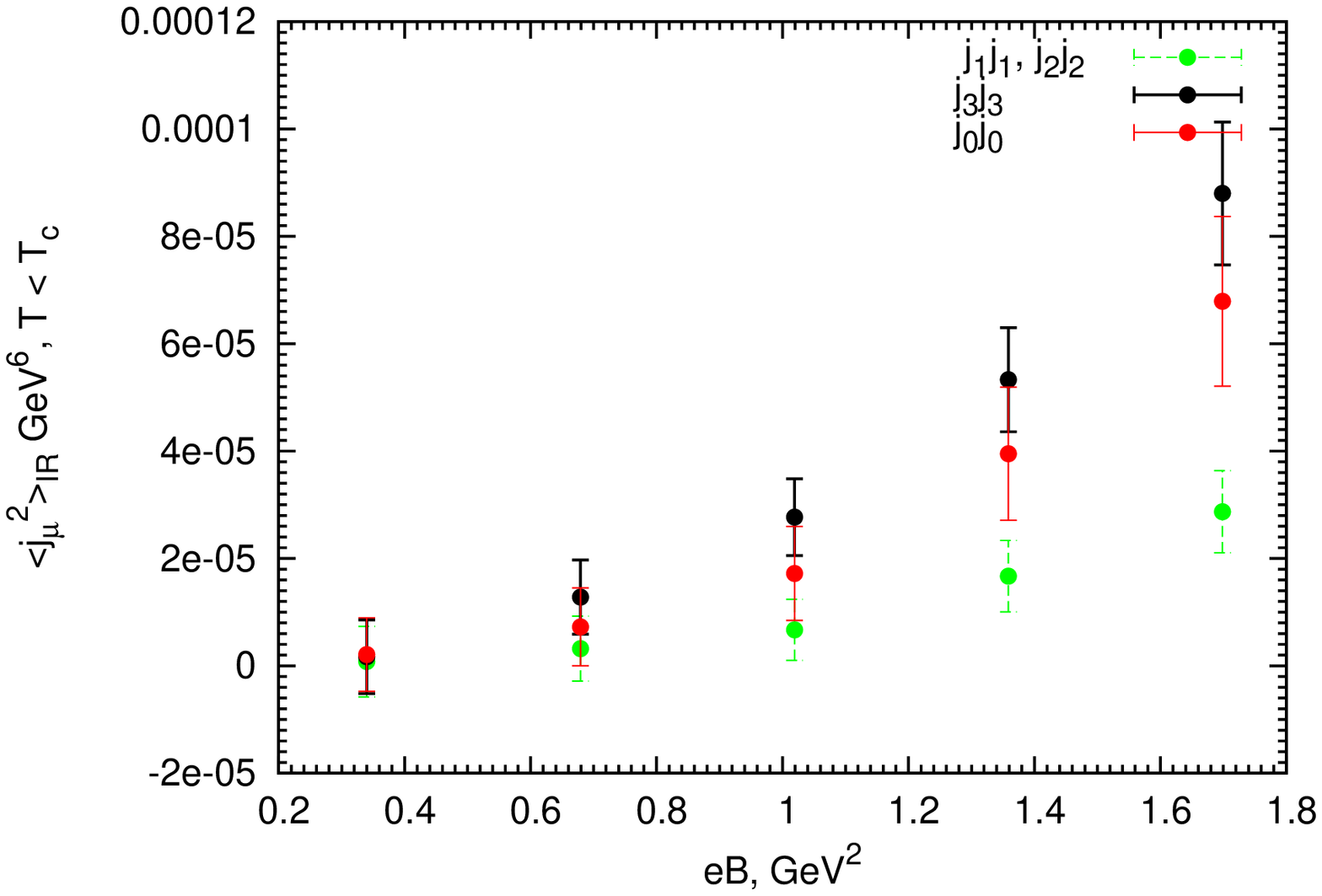}}\hspace{0cm}
\caption{\label{CMEplots} Fluctuations of the chirality (a) and electromagnetic
current/charge (b).}
\end{figure}

One example of a new effect mentioned in the Introduction is
the chiral magnetic effect (CME), which generates an electric current along the
magnetic field in the presence of a nontrivial gluonic background \cite{Kharzeev, CME}.
This effect may naturally take place in heavy-ion collisions and is at the moment under active experimental search \cite{STAR,PHENIX,ALICE} (see also a review \cite{Bzdak:2012ia} on the interpretation of the experimental data). Lattice evidences of the
effect can be found in \cite{Buividovich_CME, conductivity, Buividovich_recent, Yamamoto}.
Here we implement the procedure from \cite{Buividovich_CME} for the $SU(3)$ case and study the local
chirality
\begin{align}
 \rho_5(x) = \bar\Psi(x) \gamma_5 \Psi(x) \equiv \rho_L(x) - \rho_R(x) \label{chirality}
\end{align}
and the electromagnetic current
\begin{align}
 j_{\mu}(x) = \bar\Psi(x) \gamma_{\mu} \Psi(x).
\end{align}
The expectation value of the first quantity can be computed by (\ref{single})
with $\mathcal{O} = \gamma_5$ and with $\mathcal{O} = \gamma_{\mu}$ for the
second quantity. Both VEV's are zero, as expected, but the corresponding
fluctuations obtained from (\ref{double}) are finite and grow with the field
strength (see Fig.~\ref{CMEplots}).
Here we use the ``IR'' subscript to emphasize, that we subtract from the quantity its
value at $B=0$:
\begin{align}
\langle Y \rangle_{IR}(B) = \frac{1}{V}\int\limits_V d^4 x\langle Y(x)
\rangle_B - \frac{1}{V}\int\limits_V d^4 x\langle Y(x) \rangle_{B=0}
\end{align}
One can interpret the enhancement of the current fluctuations either as short-living quantum fluctuations or as a charge flow.
We measured the conductivity of the vacuum to argue in support of the latter.

\section{Conductivity of the QCD vacuum}\label{second}
\begin{figure*}[t]
     \centering
     \subfigure[\label{sigma_all}]
     {\includegraphics[angle=-90, width=7.5cm]{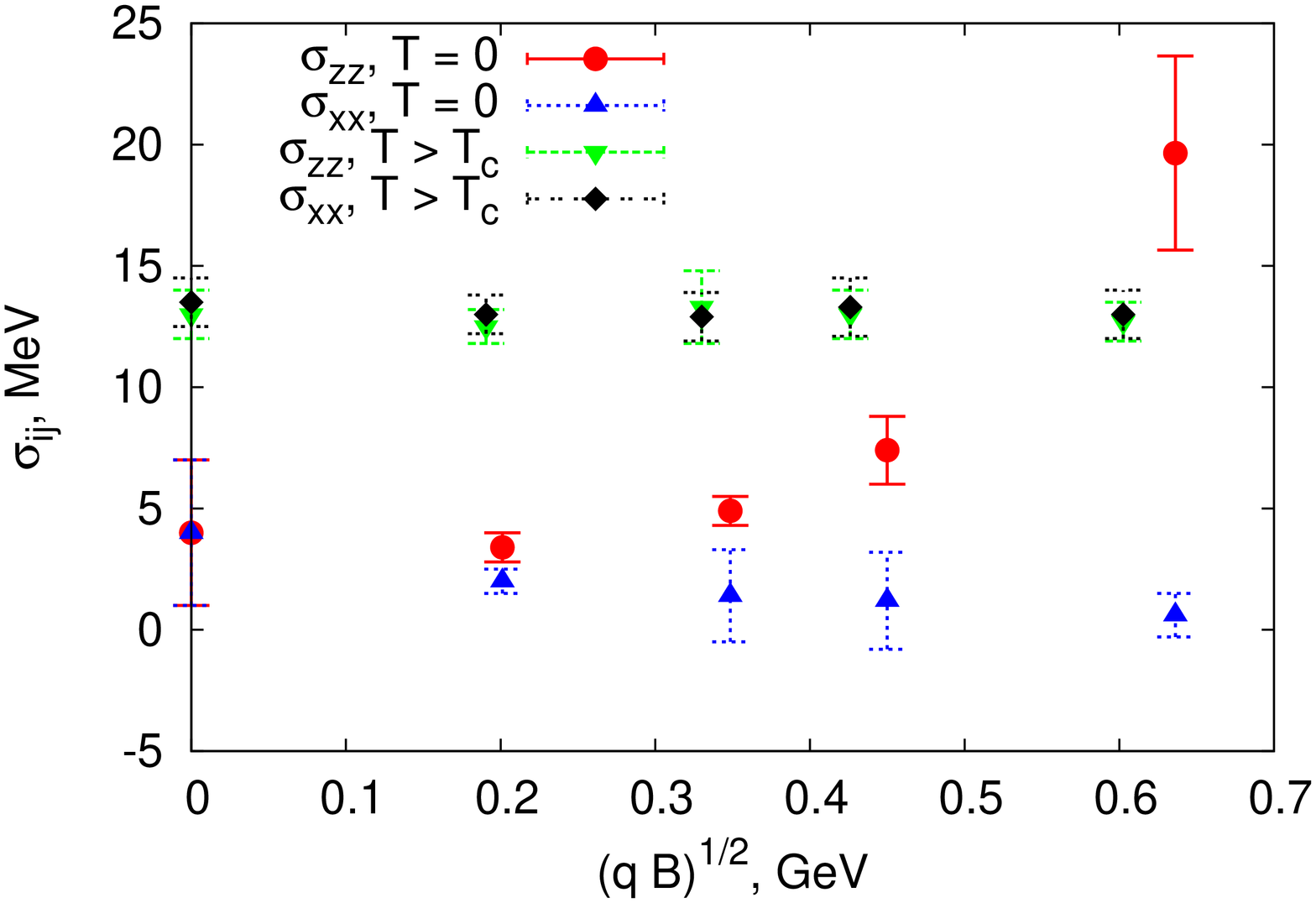}}\hspace{0cm}
     \centering
     \subfigure[\label{sigma_zz}]
     {\includegraphics[angle=-90, width=7.5cm]{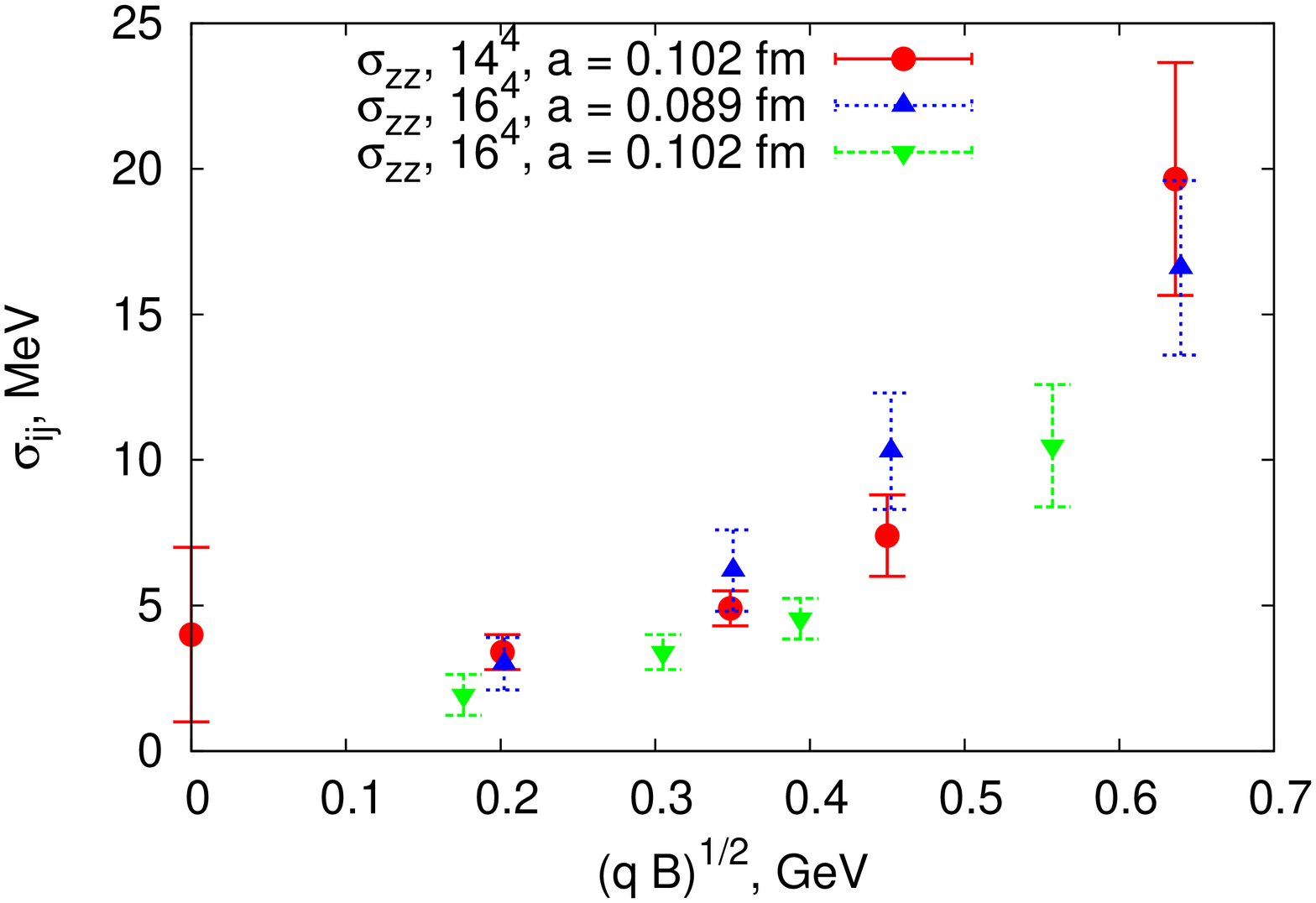}}\hspace{0cm}
  \caption{Conductivity at various temperatures (a) and at small temperatures and various lattices (b). Here $\sigma_{zz}$ ($\sigma_{xx}$, $\sigma_{yy}$) \
is the conductivity along (transverse to) the magnetic field.}
  \label{fig:conductivity}
\end{figure*}

\noindent Electric conductivity can be extracted from the correlator of two vector currents,
\begin{eqnarray}
G_{ij}\lr{\tau} = \int d^3 \vec{x} \vev{ j_{i}(\vec{0}, 0) j_{j}\lr{\vec{x}, \tau}  }
\label{corr_def}
\end{eqnarray}
Following \cite{Aarts:07:1}, let us define the spectral function $\rho\lr{w}$ which corresponds to the correlator (\ref{corr_def})
\begin{align}
\label{spectral_def}
 G_{ij}\lr{\tau} = \int\limits_{0}^{+\infty} \frac{d w}{2 \pi}\, K\lr{w, \tau} \rho_{ij}\lr{w},\quad \mathrm{with} \quad K\lr{w, \tau} = \frac{w}{2 T}  \, \frac{\cosh{\lr{w \lr{\tau - \frac{1}{2 T}}}}}
{\sinh{\lr{\frac{w}{2 T}}}},
\end{align}
where $T$ is the temperature. The Kubo formula for the electric conductivity then reads \cite{Kadanoff:63:1, Aarts:07:1}
\begin{eqnarray}
\label{Kubo_formula}
\sigma_{ij} = \lim_{\omega \to 0} \frac{\rho_{ij}\lr{\omega}}{4 T}\,.
\end{eqnarray}
In the limit of the weak time-independent electric field $E_{k}$, one has $\vev{j_{i}} = \sigma_{ik} E_{k}$. We performed the SU(2) quenched lattice computations (for details see \cite{conductivity})
and obtained results shown in Fig.~\ref{fig:conductivity}.
One can conclude from the results, that the QCD vacuum is an isotropic conductor in the deconfinement phase, and the conductivity does not depend on the strength of the
magnetic field ($\sigma = 15\pm2\, \mathrm{MeV}$ for $T=350\,\mathrm{MeV} > T_c$). In the confinement phase the conductivity in the transverse directions is zero within error range, while it is a growing function of B-field in the longitudinal direction. In other words, the strong magnetic field at small temperatures turns the QCD vacuum from an insulator to an anisotropic conductor.

Such a behavior may also lead to the enhancement in the soft photon and dilepton production rates in the direction {\it transverse} to the magnetic field \cite{Buividovich_recent}.

\section{Fractal dimension of the chirality distribution}\label{third}

\begin{figure*}[t]
\subfigure{\includegraphics[angle=-90, width=7cm]{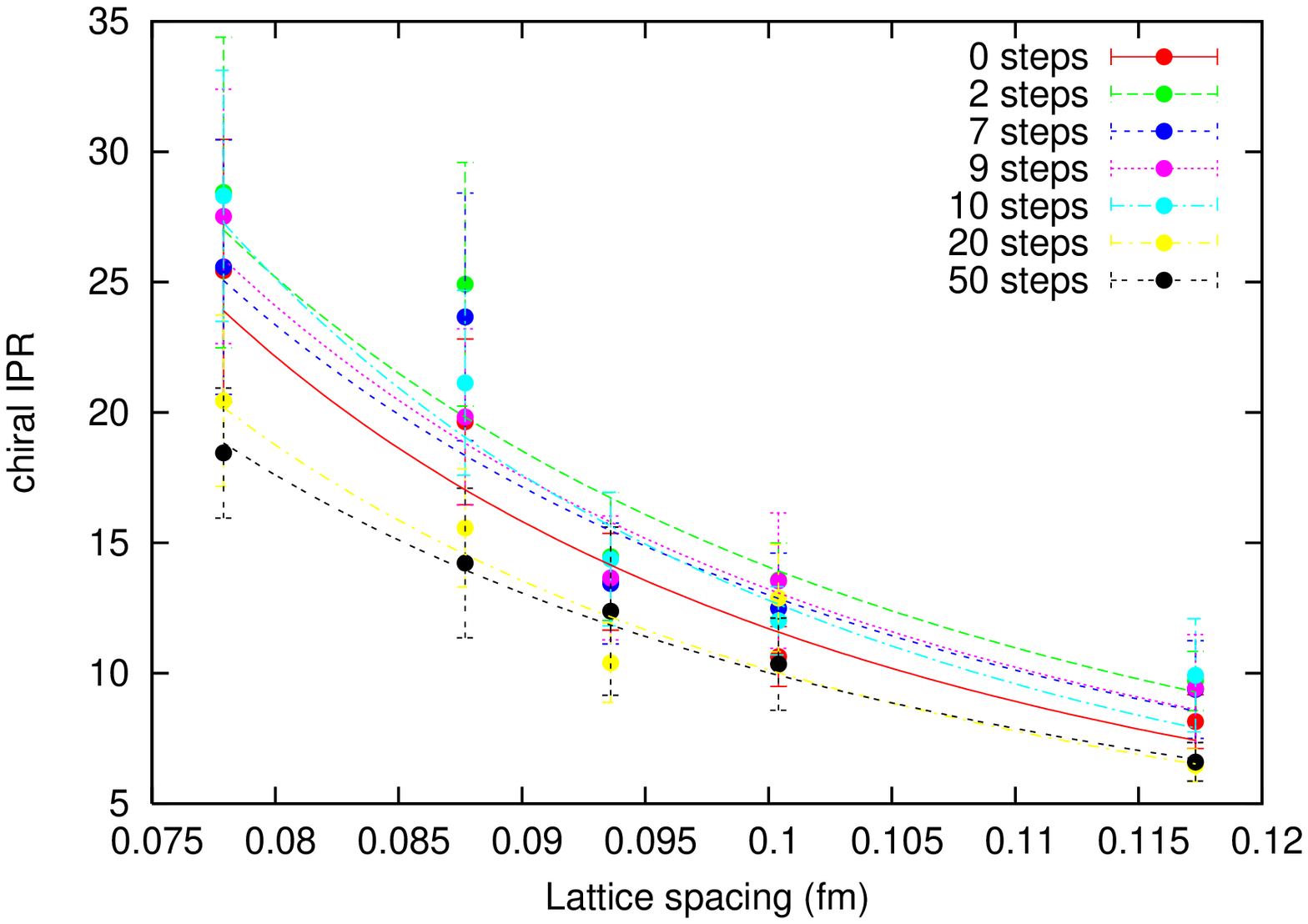}}\hspace{1cm}
\centering
\subfigure{\includegraphics[angle=-90, width=7cm]{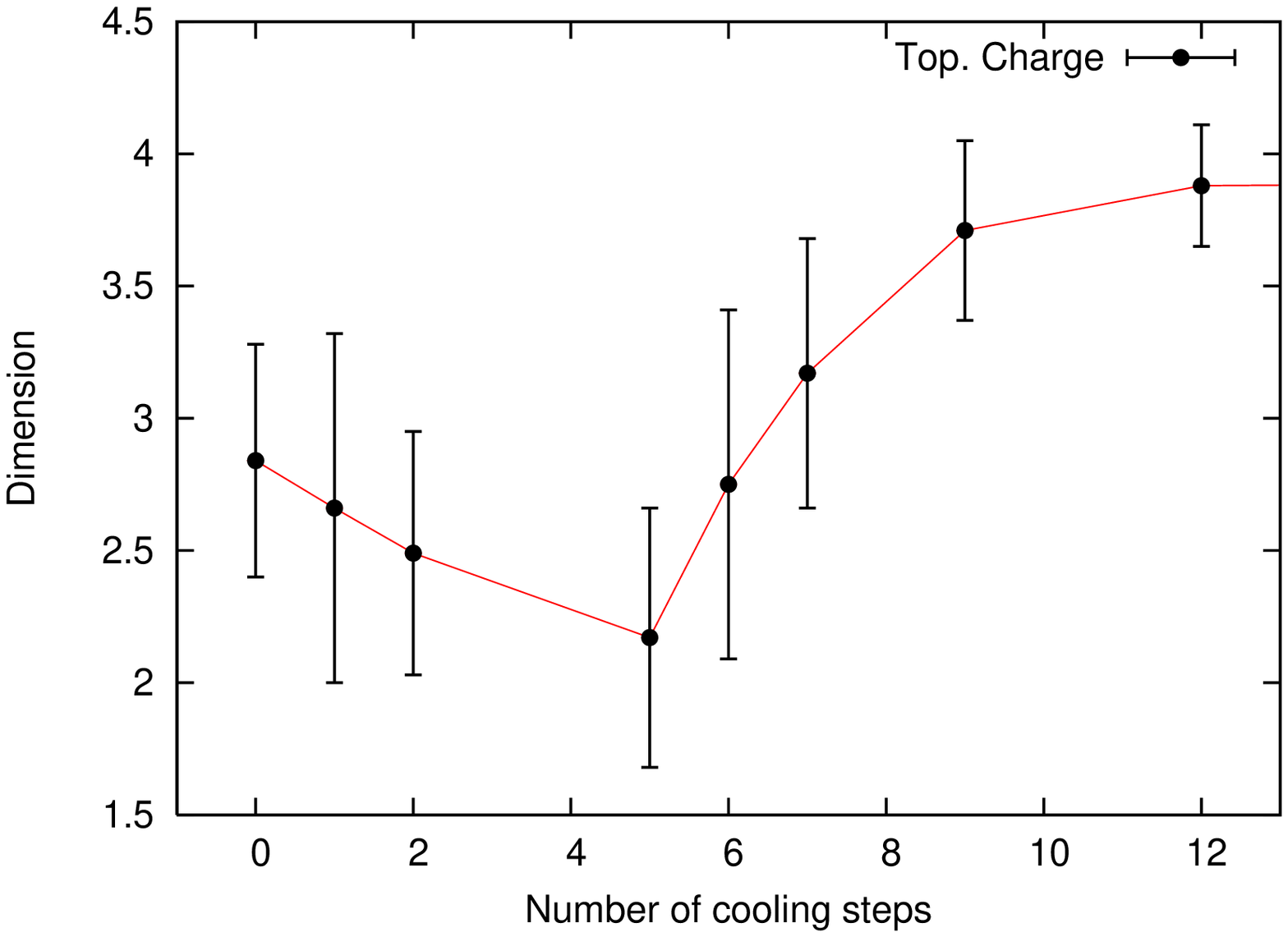}}
\vspace{-0.3cm}
\caption{Chiral IPR (defined in the text) as a function of lattice spacing (a), and the dimensionality of the chirality distribution at various cooling stages (b).}\label{aubin_ipr}
\end{figure*}

We studied some properties of the spatial distribution of chirality (\ref{chirality}), such that the volume occupied by the distribution and the fractal dimension of that volume,
for various lattice spacings and cooling stages. To measure the former we used the inverse participation ratio (IPR),
\begin{align}
\mathrm{IPR^5_0} = N \left[\ddd\frac{\ddd\sum\limits_x  \left|\rho_5(x)\right|^2}{\left(\ddd\sum\limits_x \, |\rho_5(x)|\right)^2}\right]_{\lambda=0} \equiv \frac{V_{lattice}}{V_{distribution}}\,,\label{DEF4}
\end{align}
where $N$ is the total number of sites of the lattice and the brackets $[...]_{\lambda=0}$ denote an averaging over all zero modes and further averaging over all gauge field configurations. 
In general, the IPR is equal to the inverse fraction of sites occupied by the support of a distribution.
The fractal (Hausdorff) dimension $d$ can be extracted from the IPRs at various lattice spacings $a$,
\begin{align}
 \mathrm{IPR}^5_0(a) = \ddd\frac{c}{a^{d}}\label{fit}\,.
\end{align}
Lattice parameters and simulation details can be found in \cite{Buividovich:2011cv}. The results are shown in Fig.~\ref{aubin_ipr}. Our conclusion is that the chirality distribution
(and hence fermionic zero modes) has zero volume in the continuum limit (since IPR grows with decreasing $a$) and its fractal dimension equals $d=2..3$ for uncooled configurations (in agreement with \cite{Horvath, Kovalenko:2005rz, Ilgenfritz:2007xu, Ilgenfritz:2007b}),
tending to $d=4$ after the cooling. This behavior supports the idea of \cite{Zakharov:2006te}, stating that the Yang-Mills vacuum structure is different for different resolutions of the measuring procedure.
The low dimensional structure of the vacuum, if true beyond the probe quark limit, may lead to a new phenomenology relevant for the heavy-ion experiments \cite{Kalaydzhyan:2012ut, Zakharov:2012vv, Chernodub:2012mu,Zhitnitsky:2012ej}.

\end{document}